\documentclass[conference]{IEEEtran}
\IEEEoverridecommandlockouts
\usepackage{cite}
\usepackage{amsmath,amssymb,amsfonts}
\usepackage{graphicx}
\usepackage{CJKutf8}
\usepackage{textcomp}
\usepackage{xcolor}
\usepackage{algorithm} 
\usepackage{algpseudocode} 
\usepackage{comment}
\usepackage{subcaption}
\newcommand*{\note}[1]{\textcolor{black}{#1}}
\usepackage{mathrsfs}
\usepackage{geometry}
\renewcommand{\baselinestretch}{0.98}

\def\BibTeX{{\rm B\kern-.05em{\sc i\kern-.025em b}\kern-.08em
    T\kern-.1667em\lower.7ex\hbox{E}\kern-.125emX}}
\begin{document}
\begin{CJK}{UTF8}{gbsn}

\title{Collaborative Satellite Computing through Adaptive DNN Task Splitting and Offloading}

%Multi-Satellite Collaborative Satellite Computing Empowered by DNN-based Task Partitioning and Offloading
%Multi-Satellite Collaborative Computing With DNN Partitioning and Task Allocation 
%Joint Multi User DNN Partitioning and Multi Satellite Collaborative Computing Strategy
% {\footnotesize \textsuperscript{*}Note: Sub-titles are not captured in Xplore and should not be used}

\author{\IEEEauthorblockN{Shifeng Peng$^{1}$, Xuefeng Hou$^{1}$\thanks{The first two authors contributed equally to this work.
}, Zhishu Shen$^{1}$\textbf{\IEEEauthorrefmark{2}}\thanks{\IEEEauthorrefmark{2} Corresponding author.}, Qiushi Zheng$^{2}$, Jiong Jin$^{2}$, Atsushi Tagami$^{3}$, and Jingling Yuan$^{1}$}

%\author{\IEEEauthorblockN{Xuefeng Hou$^{1}$, Shifeng Peng$^{1}$\thanks{The first two authors contributed equally to this work.}, Zhishu Shen\textbf{\textsuperscript{**}}$^{1}$\thanks{\textbf{**} Corresponding author.}, Qiushi Zheng$^{2}$, Jiong Jin$^{2}$, and Atsushi Tagami$^{3}$}

%\author{\IEEEauthorblockN{Zhishu Shen$^{1}$, Qiushi Zheng$^{2}$, Jiong Jin$^{2}$, and Atsushi Tagami$^{1}$ }
\IEEEauthorblockA{\textsuperscript{$^{1}$}School of Computer Science and Artificial Intelligence, Wuhan University of Technology, China\\
%\{\note{XX}，psf\}@whut.edu.cn, z\_shen@ieee.org\\
}
\IEEEauthorblockA{\textsuperscript{$^{2}$}School of Science, Computing and Engineering Technologies, Swinburne University of Technology, Australia\\%John Street, Hawthorn, Victoria, 3122, Australia.\\
%\{qiushizheng, jiongjin\}@swin.edu.au\\
}
\IEEEauthorblockA{\textsuperscript{$^{3}$}KDDI Research, Inc., Japan\\ %2-1-15 Ohara, Fujimino-shi, Saitama, 356-8502, Japan.\\
%at-tagami@kddi.com
}
\IEEEauthorblockA{E-mail: \{psf, martinhou, yjl\}@whut.edu.cn, z\_shen@ieee.org, \{qiushizheng, jiongjin\}@swin.edu.au, at-tagami@kddi.com}
}

% email address or ORCI}
% \and
% \textit{name of organization (of Aff.)}}
% \and
% \textit{name of organization (of Aff.)}}
% \and
% \and
% \and

\maketitle

\begin{abstract}
 Satellite computing has emerged as a promising technology for next-generation wireless networks. This innovative technology provides data processing capabilities, which facilitates the widespread implementation of artificial intelligence (AI)-based applications, \note{especially for image processing tasks involving deep neural network (DNN)}. With the limited computing resources of an individual satellite, independently handling DNN tasks generated by diverse user equipments (UEs) becomes a significant challenge. One viable solution is dividing a DNN task into multiple subtasks and subsequently distributing them across multiple satellites for collaborative computing. However, it is challenging to partition DNN appropriately and allocate subtasks into suitable satellites while ensuring load balancing. To this end, we propose a collaborative satellite computing system designed to improve task processing efficiency in satellite networks. Based on this system, a workload-balanced adaptive task splitting scheme is developed to equitably distribute the workload of DNN slices \note{for collaborative inference}, consequently enhancing the utilization of satellite computing resources. Additionally, a self-adaptive task offloading scheme based on a genetic algorithm (GA) is introduced to determine optimal offloading decisions  within dynamic network environments. The numerical results illustrate that our proposal can outperform comparable methods in terms of task completion rate, delay, and resource utilization.

%shows that under the conditions of acceptable maximum task processing time and completion rate, our algorithm has lower latency and higher task completion rate compared to the other comparable methods.

\end{abstract}


\section{Introduction}
The beyond 5G or 6G (B5G/6G) networks are expected to meet the ubiquitous demands of the future digital society. However, the traditional base stations (BSs) are unrealistic to be fully deployed in remote rural areas due to geographical and economic restrictions. Furthermore, the proliferation of mobile devices and the escalating demand for smart applications have given rise to a surge in computationally intensive tasks. In the absence of robust ground network infrastructure, processing these tasks generated by users in remote rural areas becomes a pivotal concern. Satellite computing possesses a unique strength in providing global coverage, extending connectivity to areas beyond the reach of conventional terrestrial networks. Furthermore, leveraging on-board computing capabilities in satellites for task processing can notably reduce the service delay caused by the network congestion~\cite{shen2023survey, WangIoT23}.

% Continuous interaction with the central server is required when processing computing tasks.
% Current works\cite{vasisht2021l2d2},\cite{deng2019next} adopt a centralized architecture for satellite task offloading which 
Task offloading within a centralized architecture results in substantial communication and computational overhead, particularly within large-scale low Earth orbit (LEO) satellite networks. To solve this issue, a distributed architecture that relies on data sharing among various satellites is anticipated. In this system, each terminal such as a satellite, independently determines offloading decisions based on its local observations\cite{r2}. However, the distributed solutions encounter non-convergence issues in the pursuit of global optimization. % in task offloading. 

%learning from nodes with only local observations is challenging in distributed architectures. Especially the distributed solutions encounter non-convergence issues when it comes to achieving global optimization.%to achieve a satisfactory performance.%\cite{mai2021cloud} that obtain the optimal solution in the entire search space.

%To deal with various task processing services efficiently such as object recognition and anomaly detection 
Deep learning methods based on deep neural networks (DNN)  have been widely adopted for the efficient processing of diverse tasks within satellite networks. However, the challenge arises from the limited computing and storage resources available on satellites when executing computationally intensive tasks using DNN~\cite{zhang2023satellite}. To address this limitation, large DNN tasks can be initially distributed into different \textit{blocks}  according to the task processing units determined by the decision-making satellite. These blocks are divided into multiple \textit{segments} based on network information including network scale and resource usage status. In this paper, the above procedure is referred to \textit{task splitting}. Furthermore, to enhance feature representation, DNN models frequently incorporate a substantial number of layers, thereby intensifying the complexity of \textit{DNN partitioning}. Specifically, DNN partitioning encounters the challenge of balancing the partition granularity and workload (the calculation amount of each task segment)~\cite{r5}. Finally, multiple satellites collaborate to execute these tasks, ensuring the optimal resource utilization.% of available resources. %\note{This satellite computing process can be implemented on Raspberry Pi.} 

Our primary focus is on situations in which data services encompass multiple satellites and user equipments (UEs) distributed across varied geographical regions. Diverse UEs generate a range of tasks and offload split task segments to multiple satellites for collaborative processing. %To achieve load balancing among various satellites, task splitting and offloading need to be jointly optimized. 
In this paper, we propose adaptive DNN task splitting and offloading schemes within collaborative satellite computing systems, aiming to attain load balancing across different satellites. To this end, a genetic algorithm (GA) is used to explore optimal solutions in a large decision space. The primary contributions of our work include: %are summarized as follows:

\begin{enumerate}
\item A system model is developed to facilitate collaborative computing among multiple satellites. This model encompasses DNN partitioning and task offloading, focusing on optimizing task completion rate and minimizing processing delay.

\item \note{A scheme leverages binary monotonicity is introduced for achieving workload balance by task splitting. This scheme employs dichotomy to balance the workload of DNN slices and then distributes them to multiple LEO satellites, enabling collaborative inference and enhancing the utilization of satellite computing resources.}

%\item \note{A scheme for achieving workload balance through task splitting is introduced, wherein the binary monotonicity is harnessed. This scheme uses dichotomy to balance the workload of DNN slices, then distributes them to multiple LEOs for collaborative inference, which enhances the utilization of satellite computing resources.}

\item \note{To facilitate collaborative satellite computing, a GA-based self-adaptive task offloading scheme is introduced. This scheme optimizes offloading decisions by analyzing the task inference process, ensuring the efficient transmission of intermediate results from DNN slices. }

%\item \note {To facilitate collaborative satellite computing, a GA-based self-adaptive task offloading scheme is introduced. This scheme optimizes the task inference process to determine the optimal offloading decision for transmitting the intermediate results of DNN slices. }

%\item a GA-based self-adaptive task offloading scheme is introduced to determine the optimal offloading decision through task inference process and transmission of intermediate results of DNN slices. which sets a foundation for collaboration between LEOs

%\item An experimental investigation conducted on two representative DNN types underscores the superior performance of our proposed scheme compared to other comparable methods in terms of task completion rate and task delay.
\item An experimental investigation is carried out on two representative DNN types. \note{The entire process can be executed on Raspberry Pi, a single-board computer installed on the recently launched satellite research platform, which supports in-orbit computing~\cite{WangIoT23}}. This investigation highlights the superior performance of our proposed scheme in comparison to other methods, specifically in terms of task completion rate and task delay.

%The simulation results conducted on two types of typical DNNs show that our proposal can outperform other comparative methods in terms of various criteria, i.e., by achieving at least 0.14s reduction in service delay and 4\% improvement in task completion rate. %, e.g., our scheme reduces at least 140 mseconds in service delay and makes a 4\% improvement in task completion rate.} %To test the performance of the scheme, ResNet101 and VGG19 are adopted to validate the effectiveness of our algorithm. Results show that the total average delay was reduced by 0.62s and 0.14s compared to Residual-Resource-Priority (RRP) and DQN, and the task completion rate was improved by 4\% compared to Random, RRP and DQN.
\end{enumerate}

The remainder of this paper is organized as follows: Section II summarizes the related work, and Section III describes the problem statement. Section IV presents our proposed task splitting and offloading schemes. Section V summarizes the evaluation results that verify the performance of our proposal against that of the comparable methods. Section VI gives the conclusions for our future work.

%\vspace{-0.2cm}
\section{Related Work}

\subsection{DNN Partitioning and Task Splitting}

% Most existing works design DNN partitioning algorithms with the goal of reducing latency and energy consumption. In \cite{r6}, the authors investigated the operating state of the system under light and heavy load conditions. Based on this, DNN Surgery Light(DSL) which can minimize the overall delay to process one frame was developed. The DSL algorithm utilizes the minimum s-t cut theory to minimize task processing latency. \cite{r7} proposed a reliable DNN partitioning strategy based on DQN. This algorithm updates the network with drone status, offloading location, and environmental reward values to determine the optimal offloading location for the DNN layer.
The characteristics of DNN models enable them to be easily partitioned and deployed to multiple satellites for collaborative computing. In general, DNN model partitioning can be divided into vertical partitioning~\cite{zhang2020adaptive}, horizontal partitioning~\cite{li2020edge}, and combined vertical-horizontal~\cite{zhang2021deepslicing} partitioning. In~\cite{li2020edge}, the authors integrated BranchyNet with DNN models and utilized an early exit mechanism to divide the model in order to meet the requirements of inference latency and accuracy. In~\cite{zhao2018deepthings}, the authors divided the convolutional layers based on the grid and allocated them to multiple devices with different computing capabilities for task execution. It is essential to consider the equitable distribution of workload for each segment during the DNN partitioning process to optimize resource utilization.% in satellite networks.

%To improve the resource utilization of satellite networks, the balance of the workload of each segment needs to be further considered during the DNN partitioning. 

%In the above studies, whether the workload of each block is balanced after DNN partitioning has not been considered. Balancing the workload of each piece of work can maximize the utilization of resources of the equipment participating in the task execution. Based on this we propose a BP algorithm which can balance the workload for DNN partitioning.

%\vspace{-0.1cm}

\subsection{Task Offloading in Satellite Networks}
% In \cite{r8}, a stochastic computation offloading problem was proposed for LEO satellite edge computing networks. The authors proposed a algorithm combining deep reinforcement learning (DRL) and heuristic algorithms to help users and the LEO satellite make computation offloading and resource allocation decisions.

%Regarding the recent work utilizing satellite networks for computing task offloading, 
The authors in \cite{r2} proposed a multi-agent double actors twin delayed deterministic policy gradient (MA-DATD3) algorithm that contains double actors and double critics. MA-DATD3 was designed to solve the computation offloading optimization problem with a centralized training and decentralized execution paradigm. In \cite{cheng2022dynamic}, the author conducted a joint optimization of computation offloading and power control. They introduced a Lyapunov optimization-based algorithm for minimizing the overall delay of tasks while satisfying the energy constraints of satellites. In \cite{tang2021computation}, the authors transformed the offloading decision optimization problem into a linear programming problem by using the binary variables relaxation method. A distributed algorithm based on the alternating direction method of multipliers (ADMMs) was proposed to approximate the optimal solution with low computational complexity. The methods introduced in previous research primarily catered to scenarios with a limited number of satellites covering a single area, rendering them unsuitable for offloading extensive DNN tasks that demand substantial computing resources.

%The methods proposed in the previous work mainly designed for that deploys a limited number of satellite covering a single area, which is not applicable for offloading massive DNN tasks requiring large amounts of computing resources. %To address this issue, we propose a multi-satellite collaborative task offloading scheme based on GA.

% Qin et al. \cite{r9} studied the characteristics of computation offloading under the LEO network and proposed a joint three-tier computation offloading framework, including offloading location selection and transmission energy allocation issues. A multi-agent deep reinforcement learning (DRL) algorithm was formulated to minimize task delay and reduce energy consumption.

% In the above studies, the DRL algorithm that discretizes continuous actions was used to make offloading decisions. \note{Due to the poor parallelism of the DRL algorithm, it cannot handle multiple individuals well.} And the offloading decision-making \note{effect is not good} when constrained by constraints such as resource constraints and latency requirements. Based on this, we propose a novel genetic algorithm which can balance parallelism and constraint conditions.

%In the above studies, these proposed algorithms only consider task offloading. 

\section{Problem Statement}
\subsection{System Model}
\begin{figure}[tb!]
    \centering
    \includegraphics[width=0.98\linewidth]{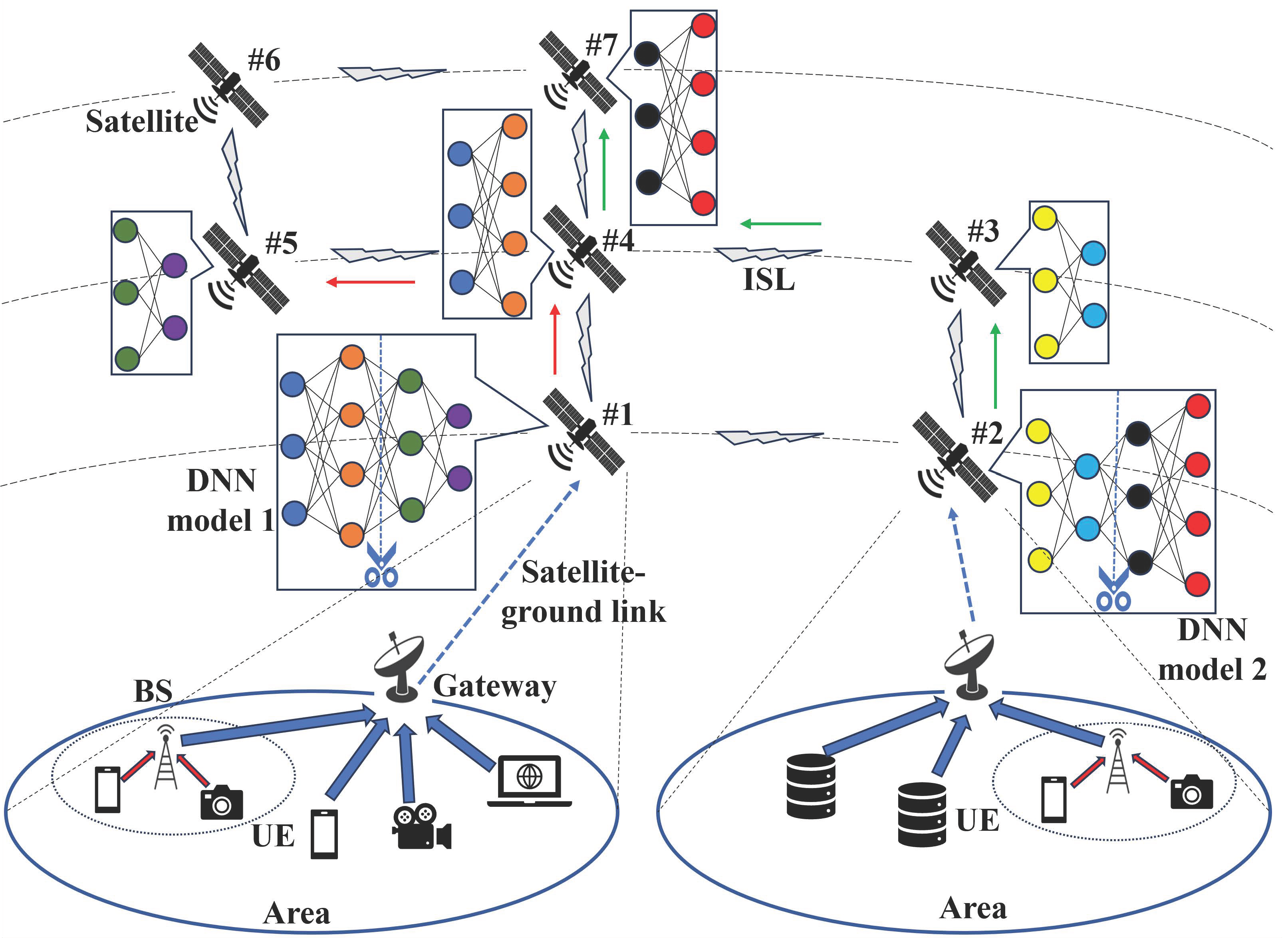}
    \caption{System model.}
    \label{fig:model}
\end{figure}

As shown in \figurename~\ref{fig:model}, we assume an integrated satellite-ground system consisting of LEO satellites and UEs located in multiple remote rural areas. Multiple LEO satellites form a constellation to achieve seamless global coverage. Each satellite orbits the Earth periodically to enable the establishment of satellite-ground connections. Due to the limited computational capabilities of UEs, these UEs have a restricted capacity to handle computationally intensive tasks. As a result, the computing tasks need to be transmitted to satellites via a gateway for processing. The gateway in each area is responsible for collecting tasks generated by the UEs within that specific area and transmitting these tasks to the satellite through wireless links. Each satellite is equipped with pre-trained DNN models to handle computing tasks collected from gateways at the ground. Moreover, satellite-to-satellite communication is facilitated through inter-satellite link (ISL). For example, in \figurename~\ref{fig:model}, satellite $\#1$ receives tasks to be conducted by DNN model $1$ from an area and executes task splitting. The segments of the respective tasks are offloaded to satellite $\#4$ and $\#5$. %through ISL. %\note{ $\#1$ is used in both satellite and DNN model}

% \begin{figure}[tb!]

% $A$ represents a set of areas, $a$ represents one of the areas.
The key elements within the proposed system comprise satellites, gateways, and UEs. Here, $\mathcal{S}$ represents a set of satellites where $i$ ($i\in \mathcal{S}$) denotes an individual satellite. Similarly, $\mathcal{G}$ stands for a set of gateways, with each gateway denoted as $g$. $\mathcal{E}$ represents a set of UEs, $E_i$ represents a set of UEs within the coverage area of satellite $i$. %$e_{i,j}$ ($e_{i,j}\in \mathcal{E})$ denotes a UEs. 
Without loss of generality, we designate the decision-making satellite to provide access and computing services to the UEs and gateway $g$ within its covered area. The system operational time is divided into $\Gamma$ slots, denoted by $\tau$. We assume that the number of computing tasks received by each decision-making satellite in slot $\tau$ obeys the Poisson distribution. Subsequently, the decision-making satellite splits the received DNN tasks and offloads the task segments to multiple satellites for collaborative computing. %Final task offload decision includes satellite sequence, e.g., $\{1, 3, 5\}$ represents that the task slices are offloaded to satellites $\#1$, $\#3$ and $\#5$ for execution respectively.
\subsection{Communication Model}
% 整个框架中涉及到的所有传输过程，地面通信点到卫星及卫星和卫星之间，这两部分并不相同。

%地面通信点到卫星%
Before the satellite determines task splitting and offloading decisions, the tasks must be transmitted to the satellite through the satellite-ground wireless link. Assuming multiple gateways share bandwidth without interfering with each other, according to Shannon formula, the average transmission rate $v_{g,i}(t)$ between gateway $g$ and satellite $i$ is: %calculated as:
\begin{equation}
   v_{g,i}(t)=B_0\log_{2}{(1+\frac{P_g\xi_{g,i}(t)}{M_G})},
\end{equation}
where $B_0$ is the channel bandwidth, $P_g$ represents the transmit power, $\xi_{g,i}(t)$ denotes the channel gain consists of large-scale fading and shadowed-Rician fading\cite{tang2021computation}, and $M_G$ indicates the additive white Gaussian noise.

%卫星与卫星之间%
The satellite network consists of $N_o$ orbits, on which $N_s$ satellites are evenly distributed. Due to the constraints imposed by communication distances, each satellite can only transmit tasks to its adjacent satellites through ISL. Assuming Gaussian channels, the maximum achievable data rate\cite{mayorga2021inter} for the transmission between satellites $i$ and $j$ is:
\begin{comment}
\begin{equation}
    r(i,j)=B\log_{2}{(1+\textit{SNR}(i,j))},
\end{equation}
\end{comment}
\begin{equation}
    r(i,j)=B\log_{2}{(1+\frac{P_tG_i(j)G_j(i)L_i(j)L_j(i)}{kTB})},
\end{equation}
where $B$ is the bandwidth between satellites, $P_t$ is the transmission power, $G_i(j)$ and $G_j(i)$ are the gain of the transmit and received antennas, respectively. $L_i(j),L_j(i)$ represent the beam pointing coefficient ($L_i(j),L_j(i)<1$), $k$ is the wavenumber, and $T$ is the resultant noise temperature.

\begin{comment}
where $B$ is the bandwidth between satellites, and $SNR(i,j)$ is the signal-to-noise ratio (SNR) which can be defined as:

\begin{equation}
    \textit{SNR}(i,j)=\frac{P_tG_i(j)G_j(i)L_i(j)L_j(i)}{kTB},
\end{equation}
\end{comment}

\subsection{Computation Model}

Each satellite is equipped with a predefined DNN model that enables it to handle specific DNN task segments. Additionally, a satellite has the capability to transmit its output to adjacent satellites for further processing, such as inference and computations like pooling and convolution for the next slice. Considering the constraints of a satellite's resource capacity and the size of the DNN task segments, it is imperative to enhance the utilization of each satellite's computing resources by achieving a balanced workload distribution for each DNN task segment. We aim to optimize the largest workload of all DNN task segments in a min-max fashion. The utility function is defined as:
\begin{equation}
    U: \min_{}\max_{k\in\{1,2,...,L\}}m_k,
\end{equation}
where $m_k$ represents the workload of each DNN task segment, $L$ denotes the expected sliced number.

In satellite network operations, a critical scenario may arise in which the satellite's remaining computing resources are depleted. In this case, the DNN task segment assigned to the respective satellite will not be executed, and thus the original task will be discarded. It is essential to determine whether a task can be processed or not on a satellite based on the current resource usage. After a satellite loads new DNN task segments, the total workload is:
\begin{equation}
    W=q+m_k, k\in\{1,2,...,L\},
\end{equation}
where $q$ represents the workload of DNN task segments a satellite has already loaded. The maximum workload that a satellite can accommodate is $M_w$. If $W<M_w$, load the task segments to the satellite for processing, otherwise, discard the respective task segment.

\subsection{Task Delay and Drop Model}

The task delay comprises computational delay and transmission delay. For each pre-split DNN task block, let $L$ be the number of segments and $\{q_{i,j,1},q_{i,j,2},\cdots,q_{i,j,L}\}$ be the workload of segments of \textit{j}-th block of satellite \textit{i}. Suppose the \textit{k}-th segment of this block is input to satellite $s_{i,j,k}$, the computation capability of satellite $x$ is represented as $C_{x}$, then the computation delay $t^\textit{comp}_x$ of satellite $x$ can be deduced by:
\begin{equation}
   t^\textit{comp}_x = \frac{1}{C_x}\sum \limits_{i \in \mathcal{S}} \sum \limits_{j = 1}^{N^\textit{task}_i} \sum \limits_{k=1}^L q_{i,j,k} \cdot [s_{i,j,k}=x],
\end{equation}
where $[\cdot]$ is a self-reference function defined as: %. In other words, $[x_{j,k,l}=i]= 1$  if and only if $x_{j,k,l}=i$ holds, otherwise $[x_{j,k,l}=i] = 0$. 
\begin{equation}
[s_{i,j,k}=x]=\left\{
\begin{aligned}
&1   ,  &s_{i,j,k}=x, \\
&0, &Otherwise. 
\end{aligned}
\right.
\label{eq:total_delay}
\end{equation}

For the transmission delay, let $\textit{MH}(i,j)$ be the  
Manhattan distance between satellite $i$ and satellite $j$. The transmission delay $t^\textit{tran}_x$ of satellite $x$ is calculated by:
\begin{equation}
   t^\textit{tran}_x =\sum \limits_{i \in \mathcal{S}} \sum \limits_{j = 1}^{N^\textit{task}_i} \sum \limits_{k=1}^{L-1} \textit{MH}(s_{i,j,k},s_{i,j,k+1}) \cdot q_{i,j,k}  \cdot [s_{i,j,k}=x].
\end{equation}

The total delay $t^\textit{sum}_x$ of satellite $x$ is defined as:
\begin{equation}
   t^\textit{sum}_x=t^\textit{comp}_x + t^\textit{tran}_x.
   \label{eq:delay}
\end{equation}

In addition, the task drop could happen when the computing resources of the current satellite are insufficient. Let $D_{i,j}$ be the number of drop of the $j$-th task block of satellite $i$, the total drop rate $r_D$ could be calculated as:
\begin{equation}
   r_D = \frac{\sum \limits_{i \in \mathcal{S}} \sum \limits_{j=1}^{N^\textit{task}_i} D_{i,j}}{\sum \limits_{i \in \mathcal{S}} N^\textit{task}_i}.
\end{equation}

\subsection{Optimization Objective}

Based on the aforementioned models, we aim to minimize the task drop rate and delay during the whole process using the non-negative weight parameters $\alpha$ and $\beta$. The optimization problem is defined as below: 
\begin{equation}
\min \limits_{a \in \mathcal{A}} \quad (\alpha r_D + \beta \sum \limits_{i \in \mathcal{S}}t^\textit{sum}_i), 
\label{eq:opt}
  \end{equation}
s.t.
\begin{subequations}  
\begin{equation}
 r_D \le r^\textit{max},
\end{equation}
\begin{equation}
 \sum \limits_{i \in \mathcal{S}}t^\textit{sum}_i \le t^\textit{max},
\end{equation}
\begin{equation}
  \textit{MH}(x,s_{i}) \le D_M,\forall s_{i} \in \mathcal{A}_{x},
 \label{eq:con_c}
\end{equation}
\begin{equation}
 dp_{i,j} \in \{1,2,\cdots,L+1\}, \quad \forall i \in \mathcal{S}, 1 \le \forall j \le N_i^\textit{task},
 \label{eq:con_d}
\end{equation}
\begin{equation}
 N^{l}_{i,j} \ge L,\quad \forall i \in \mathcal{S},\forall j = 1,2,\cdots,N^\textit{task}_i.  \label{eq:last_con}
\end{equation}
\end{subequations}

%\begin{equation}
%   \min (\alpha r_D + \beta \sum \limits_{i \in \mathcal{S}}t^{sum}_i), 
%\end{equation}

Regarding the constraints, the first two constraints indicate those associated with the minimum drop rate and task delay. It is not preferable to transmit tasks to satellites that are located far away. \note{For an offloading scheme that aims to choose candidate satellites $s_{i}$ from the decision space  $\mathcal{A}_{x}$ determined by satellite $x$, the constraint is expressed in Equation~\ref{eq:con_c}. $D_M$ herein is the maximum permissible communication distance.}

%On account of communication distance, it is not desirable to transmit the tasks to satellites faraway. For any offloading scheme designed to select candidate satellites $cs_{i} \in \mathcal{A}_{x}$ for task offloading decided by satellite $x$, this constraint is shown in Equation~\ref{eq:con_c}, where $D_M$ is the maximum allowed communication distance.

Besides, the arriving tasks that can be completed according to beforehand scheme $(s_{1},s_{2},\cdots,s_{L})$ or be dropped during the progress. The drop point $dp_{i,j}$ could be only selected from $\{1,2,\cdots,L \}$ (See Equation~\ref{eq:con_d}). The whole offloading process is completed if $dp_{i,j} = L + 1$. From the aspect of DNN partitioning, the number of DNN layer $N^{l}_{i,j}$ should be strictly larger than a preset partitioned number $L$ (See Equation~\ref{eq:last_con}).

\section{DNN Task Splitting and Offloading Schemes}
%Due to the offloading decisions associated with the coupling relationship among different variables, even the single-objective optimization problem is a discrete and non-convex optimization problem, which is regarded as a classical NP-hard problem~\cite{tang2021computation}.
The offloading decisions are associated with the coupling relationship among different variables. This results in the single-objective optimization problem being a discrete and non-convex optimization problem, which is regarded as a classical NP-hard problem~\cite{tang2021computation}. To address this computational challenge, we introduce adaptive DNN task splitting and offloading schemes for collaborative satellite computing. Specifically, the satellite distributes tasks into different blocks and splits them into $L$ segments. The cooperative processing sequence between satellites is determined by the task offloading scheme, enabling each satellite to execute task processing based on the computed decision.

%\note{Due to the computational intensity of our developed multi-objective optimization problem, even the single-objective optimization associated with task offloading, which has been established as an NP-hard problem in prior work~\cite{tang2021computation}.To address this computational challenge, we introduce adaptive DNN task splitting and offloading schemes for satellite collaborative computing.}

%Then task offloading scheme decides the cooperative processing sequence between satellites, by which the satellite executes task processing according to the obtained decision. 

\subsection{Workload Balanced Task Splitting Scheme}

\begin{algorithm}[!ht]
\small
    \renewcommand{\algorithmicrequire}{\textbf{Input:}}
    \renewcommand{\algorithmicensure}{\textbf{Output:}}
    \caption{Workload Balanced Task Splitting Scheme}
    \label{alg:dnn}
    \begin{algorithmic}[1]
        \Require The workload of each layer $\{w_{1},\cdots,w_{N^{l}}\}$, expected sliced number $L  (L \le N^{l})$, and precision $\epsilon$. % input
        \Ensure The partitioning result with $L$ slices. % output
    
        \Procedure{Split}{$LimitSize$} 
        \State Initialize $ Scheme  \leftarrow \emptyset, Block_{temp} \leftarrow \emptyset$
        \For {$i = 1,2,3,\cdots, N^{l} $}
            \If {$\sum \limits_{w \in Block_{temp}} w + w_i \le LimitSize$}
                    \State ${Block_{temp}} \leftarrow  Block_{temp} \cup w_i$
                \Else
                    \State $Scheme \leftarrow Block_{temp} \cup Scheme$
                    \State $Block_{temp} \leftarrow \emptyset$
                \EndIf
        \EndFor
        \State \Return {$Scheme$}
        \EndProcedure
        \State Initialize $Lower \leftarrow \max \limits_{k\in \{1,2,\cdots, N^{l}\}}w_k,Upper \leftarrow \sum \limits_{k=1}^{ N^{l}}w_k$  \label{alg:ini}
        \While {$Upper-Lower > \epsilon$} 
            \State Denote $mid \leftarrow \lfloor\frac{Lower+Upper}{2}\rfloor$
            \State Obtain $Scheme \leftarrow$ \textit{Split}$(mid)$
            \If {$|Scheme| > L  $}
                \State $Lower \leftarrow mid$
            \Else
                \State $Upper \leftarrow mid$
            \EndIf
        \EndWhile
        \State Let $result \leftarrow \textit{Split}(Upper)$
        \If {$|result| < L$}
            append empty blocks of segments to $result$ till $|result| = L$ \label{alg:append}
        \EndIf
        \State \Return $result$
\end{algorithmic}
\end{algorithm}

Achieving workload balance among resource-constrained satellites requires distributing the workload of each task block evenly. This is crucial to avoid overwhelming specific satellites with massive tasks, while others remain underutilized. For this purpose, we propose a workload balanced task splitting scheme as illustrated in Algorithm~\ref{alg:dnn}.
%In terms of workload balancing in each resource-constraint satellite, it is significant \note{emphasizing the need} to evenly distribute the workload of each task block.

\begin{comment}
\note{can't understand}Since the workload that can be accommodated by each layer of DNN is limited, the number of segments decreases with the increase of the maximum acceptable workload of a slice. %\note{Each layer has a limited workload capacity, and the number of \textcolor{blue}{segments} decreases as the maximum acceptable workload per slice increases.} 
Let $l_1,l_2$ $(l_1 \le l_2)$ be the size limitation of workload on each segment，%\note{represent the size limitations of workloads}, 
and $f(l)$ be the segments number with size limitation $l$, then $f(l_1) \ge f(l_2)$ holds, which means a minimum-maximum acceptable workload can be obtained by Binary Search.%\note{The relationship $f(l_1) \ge f(l_2)$ holds, signifying that a binary search can be deployed to determine the minimum-maximum acceptable workloads.}
\end{comment}

Initially, a lower bound (\textit{Lower}) is set as the maximum workload of each layer. This step is essential to ensure that each segment can be accurately processed (Line~\ref{alg:ini} in Algorithm~\ref{alg:dnn}). Assembly, an upper bound (\textit{Upper}) is defined as the total workload of all layers, which indicates that all layers can be partitioned into a single block.

Then, our scheme confirms whether it is appropriate to set the maximum workload of a block as $mid$ or not (See \textit{Split}() in Algorithm~\ref{alg:dnn}). If the obtained resulting number of slices is less than the partitioned number $L$, in accordance with monotonicity, the maximum acceptable workload of block needs to be decreased, otherwise, the number of blocks should be increased. After performing a Binary Search, it is possible that the block number is less than $L$. In this case, empty blocks that signify the absence of any workload are added to the final result (Line~\ref{alg:append} in Algorithm~\ref{alg:dnn}).

%empty blocks indicating no workload existed are appended to the final result (Line~\ref{alg:append} in Algorithm~\ref{alg:dnn}).

%the sake of the following offloading algorithm, we could append empty block, which has no workload, to our result.

The time complexity of Binary Search is $O(\log_2V)$, where $V$ is the value field of searching interval. In our proposed scheme, as the algorithm enumerates workloads for all layers during the split procedure, the time complexity of Algorithm~\ref{alg:dnn} is $O(N^{l} \cdot \log_2V)$. Meanwhile, in terms of space complexity, additional memory is only needed when storing the results of the split procedure. Therefore, the space complexity is $O(L)$.

\subsection{GA-based Self-adaptive Task Offloading Scheme}

\begin{algorithm}[!ht]
\small
    \renewcommand{\algorithmicrequire}{\textbf{Input:}}
    \renewcommand{\algorithmicensure}{\textbf{Output:}}
    \caption{GA-based Self-adaptive  Task Offloading Scheme}
    \label{alg:GA}
    \begin{algorithmic}[1]
        \Require The workload of sliced block $\ B_{i,j} =\{q_{i,j,1},q_{i,j,2},\cdots,q_{i,j,L}\} \in \mathcal{B}_{i}$, and the set of indices of available satellites $\mathcal{S}_{avai}$.
        \Ensure The task offloading result.
        \State Initialize primitive group $gp_0$ by randomly summoning $N_{ini}$ chromosomes like $\mathcal{C} = (c_1,c_2,\cdots,c_L)$ of length $L$, $\forall i =1,2,\cdots,L, c_i \in$ satellite indices. \label{alg:n1}
        \For{$i=1,2,3,\cdots,N_{iter}$}
            \If {($i \neq 1) \wedge (| \min \limits_{j \in gp_i} \textit{def}_j - \min \limits_{j \in gp_{i-1}} \textit{def}_j | \le \epsilon$)}
                \State break
            \EndIf
            \State Reproduce any different chromosomes $\mathcal{C}$ and $\mathcal{D}$ pairwise by heuristic algorithm \label{alg:l6}
            
            %more concretely, for 2 chromosomes like ,$\forall c_i = d_j (1\le i \le j \le L)$ summon $(d_1,d_2,\cdots,d_j,c_{i+1},c_{i+2},\cdots,c_{i+(L-j)})$ and $(d_{1+j-i},d_{2+j-i},\cdots,d_{j-1},c_{i},c_{i+1},\cdots,c_L)$
            
            \State Eliminate those chromosomes $(d_1,d_2,\cdots,d_L)$ with highest deficit calculated by Equation~\ref{eq:mh} till the size of $gp_i \le N_K$ \label{alg:l7}

            %$\theta_1 \sum \limits_{k = 1}^L \frac{q_{i,j,k}}{C_{s_{d_k}}} + \theta    _2 \sum \limits_{k = 1}^{L-1} q_{i,j,k} \cdot q(s_{d_k},s_{d_{k+1}}) + \theta_3 D_{i,j}$
            
            \State Randomly summon $N_{summ}$ new chromosomes  \label{alg:l8}
        \EndFor
        \State \Return The  chromosome with the lowest deficit
\end{algorithmic}
\end{algorithm}

We introduce GA in our scheme due to its ability to continuously explore and adapt task assignments. This capability is particularly valuable for effectively managing real-time changes and uncertainties in dynamic networks. After conducting Algorithm~\ref{alg:dnn}, the arriving tasks are split into $L$ segments and distributed into several blocks. Subsequently, there arises a need to determine a satellite processing sequence, i.e., $(c_1,c_2,\cdots,c_L)$ denotes the $i$-th segment of tasks of this block will be processed by satellite ${c_i}$, which is defined as the chromosome of individuals. Our scheme first initializes $N_{ini}$ individuals, a process outlined in Line~\ref{alg:n1} of Algorithm~\ref{alg:GA}. Then, a series of iterative steps are employed, encompassing reproduction, elimination, and augmentation operations, to derive a chromosome sequence that minimizes deficits.%, which corresponds to Line~\ref{alg:l6} to \ref{alg:l8} in Algorithm~\ref{alg:GA} respectively.

For the reproduction operation (Line~\ref{alg:l6} in Algorithm~\ref{alg:GA}), we propose a heuristic algorithm to reproduce individuals. Let $(c_1,c_2,\cdots,c_L)$ and $(d_1,d_2,\cdots,d_L)$ be the parents' chromosomes. For each pair of indices $i,j$ ($1\le i \le j\le L$, $c_i=d_j$), summon two new individuals, the chromosomes of which are $(d_1,d_2,\cdots,d_j,c_{i+1},c_{i+2},\cdots,c_{i+L-j-1},c_{i+L-j})$ and $(d_{j-(L-i)},d_{j-(L-i)+1},\cdots,d_{j-1},c_{i},c_{i+1},\cdots,c_{L-1},c_L)$.

For the elimination operation, a novel deficit measurement is introduced to discriminate the quality of each chromosome, the deficit of $(d_1,d_2,\cdots,d_L)$ is calculated by: 
\begin{equation}
   \theta_1 \sum \limits_{k = 1}^L \frac{q_{i,j,k}}{C_{{d_k}}} + \theta    _2 \sum \limits_{k = 1}^{L-1} q_{i,j,k} \cdot \textit{MH}({d_k},{d_{k+1}}) + \theta_3 D_{i,j},
   \label{eq:mh}
\end{equation}
where $\theta_1, \theta_2$ and $\theta_3$ are non-negative weight parameters.

%In order to keep the variety of chromosomes, after the operations of reproduction and elimination, an augment operation that spans some new individuals to participate in next iteration is conducted.

The augmentation operation is performed after the operations of reproduction and elimination to maintain diversity among the chromosomes. This operation introduces new individuals to participate in the next iteration. Eventually, the chromosome with the smallest deficit is selected as the processing sequence. % for task offloading.

During each iteration, it is necessary to reproduce different pairs of individuals, eliminate individuals to maintain the group size, and then summon several new individuals. The group size at the beginning of iteration is approximately $N_{summ} + N_K$ and the time complexity for conducting these 3 operations is $O((N_{summ}+N_K)^2 \cdot L)$. Therefore, the time complexity of Algorithm~\ref{alg:GA} is approximate to $O(N_{iter}\cdot(N_{summ}+N_K)^2 \cdot L)$. Meanwhile, the maximum size of the group during the iteration is approximate to the quantity of $(N_K+N_{summ})^2$. Since every individual has a chromosome of length $L$, the space complexity of Algorithm~\ref{alg:GA} is $O((N_K+N_{summ})^2 \cdot L)$.

% \addtolength{\footskip}{0.09in}
% \setlength{\footskip}{1in}

% \addtolength{\topmargin}{-0.04in}
\section{Experiments}
\subsection{Experimental Setup}
An experimental environment has been designed for LEO satellite networks of varying sizes, denoted as $N \times N$, where each network consists of $N$ orbits and each orbit contains $N$ satellites. The main parameters are summarized in Table~\ref{tab:params}. Considering the regularity of the constellation topology, we define the neighbors of each satellite as the adjacent four satellites that can directly communicate with each other. The incidence of each UE subjects to Poisson Distribution $\pi(\lambda)$, where $\lambda$ indicates the number of tasks. We verify the effectiveness of our proposed \textit{SCC} against the following methods:%three comparative methods:
\begin{itemize}
\item  \textit{Random} is a method where the candidate satellite for offloading is independently and randomly selected.% from $\mathcal{A}_x$.

\item  \textit{Residual-Resource-Priority (RRP)} is a method that selects the available satellites with the most residual computing resources to process the next segment of the tasks.
    
    %\item  \textit{DQN} is a commonly used DRL algorithm. It also tries to minimize the task drop rate and delay based on current observed network states. %It has a two-layer structure with 256x256.

\item  \textit{DQN} is a commonly used DRL algorithm. It endeavors to minimize the task drop rate and delay based on current observed network states.%~\cite{QiuTVT19}
    
\end{itemize}

\begin{table}[tb]
 \caption{Main experimental parameters}
 \label{tab:params}
 \centering
 \scalebox{0.95}{
\begin{tabular}{ll}
   \hline
   Parameter  & Value \\
   \hline \hline
   Network topology $N$ (size = $N \times N$) & 4 $\sim$ 32, \note{default value = 10}\\
   Satellite bandwidth $B$& 20~MHz~\cite{mayorga2021inter}\\
   Satellite computation capability $C_x$& 3~GHz~\cite{zhang2023satellite}\\
   Satellite transmission power $P_t$
   & 30~dBw~\cite{zhang2023satellite}\\
   Gateway bandwidth $B_0$ & 10~MHz~\cite{zhang2023satellite}\\
   Generated task incidence $\lambda$ & 4 $\sim$ 70\\
   Task splitting number $L$ & 3 (VGG19), 4 (ResNet101) \\ 
   Maximum communication distance $D_M$ & 2 (VGG19), 3 (ResNet101) \\
   %Weight parameter $\theta_1$, $\theta_2$, $\theta_3$ & 1, 20, $10^{6}$ \\
   %$N_{ini}$, $N_{iter}$, $N_{K}$, $N_{summ}$, $\epsilon$ & 20, 10, 20, 10, 1 \\
    $\theta_1$, $\theta_2$, $\theta_3$, $N_{ini}$, $N_{iter}$, $N_{K}$, $N_{summ}$, $\epsilon$ & 1, 20, $10^{6}$, 20, 10, 20, 10, 1  \\
   \hline
  \end{tabular}
  }
\end{table}
 %In addition, the expected sliced number $L$ and maximum communication distance $D_M$ are assigned with 4 and 3 respectively.

\begin{figure*}[t]
	\centering
	\begin{minipage}[b]{.5\columnwidth}
		\centering
		\includegraphics[width=\columnwidth]{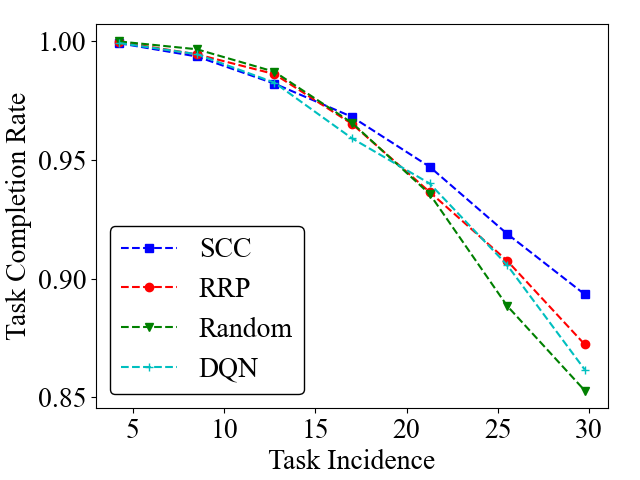}
		\subcaption{Task completion rate }\label{fig:tcr_r1}
	\end{minipage}
	\begin{minipage}[b]{.5\columnwidth}
		\centering
		\includegraphics[width=\columnwidth]{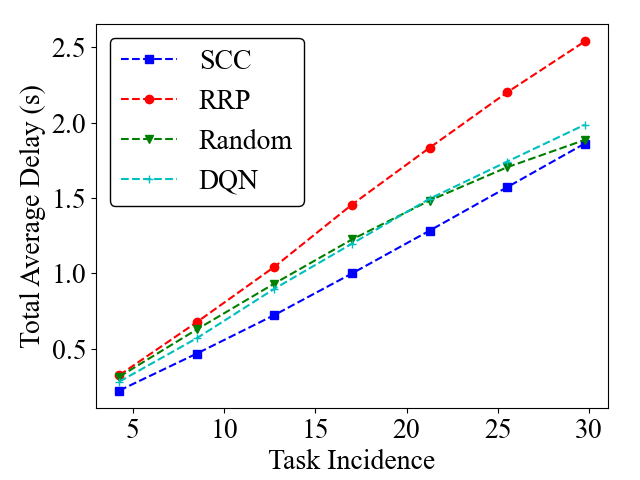}
		\subcaption{Total average delay
  }\label{fig:time_r1}
	\end{minipage}
	\begin{minipage}[b]{.5\columnwidth}
		\centering
		\includegraphics[width=\columnwidth]{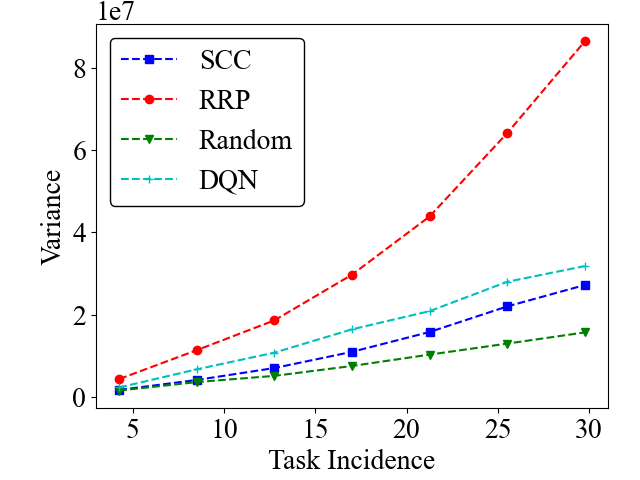}
		\subcaption{Resource usage variance}\label{fig:ru_r1}
	\end{minipage}
	\begin{minipage}[b]{.5\columnwidth}
		\centering
		\includegraphics[width=\columnwidth]{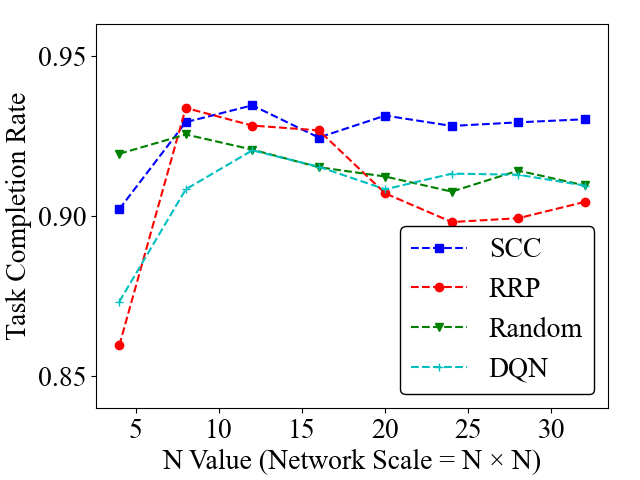}
		\subcaption{Task completion rate ($N$)}\label{fig:tcr_nw_r1}
	\end{minipage}	
	\caption{Performance achieved by different methods using ResNet101.}
	\label{fig:result1}
\end{figure*}

\begin{figure*}[t]
	\centering
	\begin{minipage}[b]{.5\columnwidth}
		\centering
		\includegraphics[width=\columnwidth]{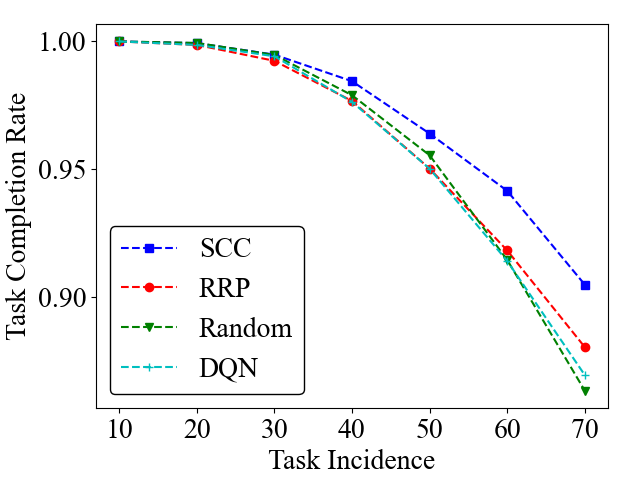}
		\subcaption{Task completion rate }\label{fig:tcr}
	\end{minipage}
	\begin{minipage}[b]{.5\columnwidth}
		\centering
		\includegraphics[width=\columnwidth]{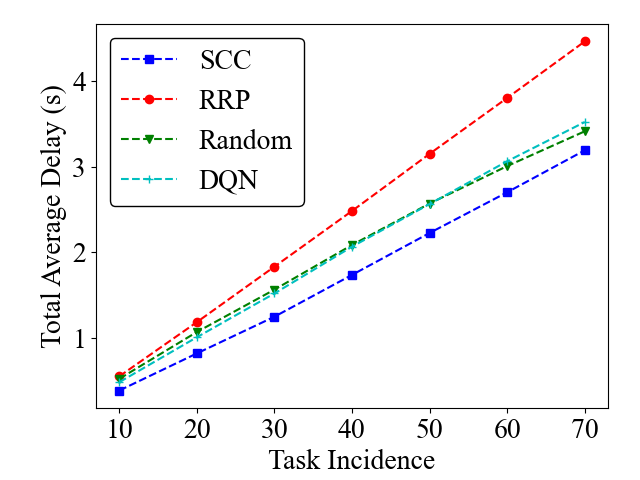}
		\subcaption{Total average delay  }\label{fig:time}
	\end{minipage}
	\begin{minipage}[b]{.5\columnwidth}
		\centering
		\includegraphics[width=\columnwidth]{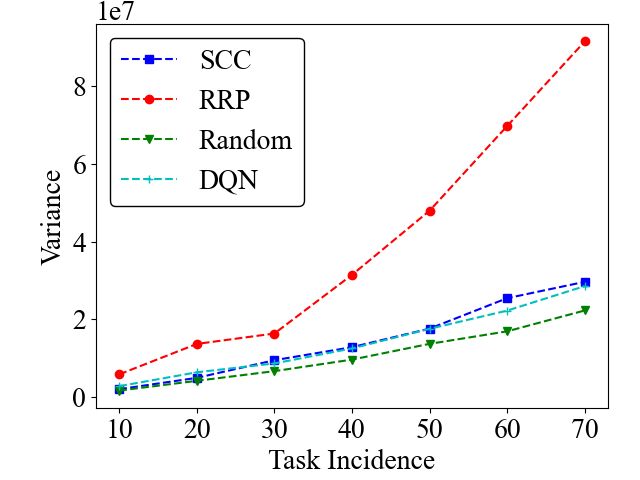}
		\subcaption{Resource usage variance}\label{fig:ru}
	\end{minipage}
	\begin{minipage}[b]{.5\columnwidth}
		\centering
		\includegraphics[width=\columnwidth]{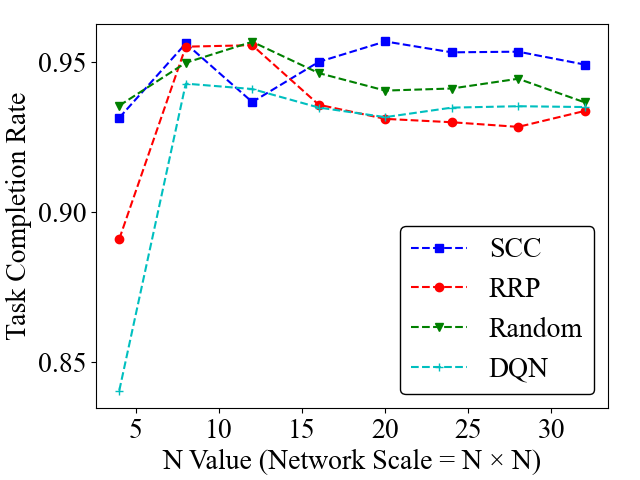}
		\subcaption{Task completion rate ($N$)}\label{fig:tcr_nw}
	\end{minipage}	
	\caption{Performance achieved by different methods using VGG19.}
	\label{fig:result2}
\end{figure*}

\subsection{Experimental Results}

%We select 2 commonly used DNN models: ResNet101 and VGG19 as our tasks, the experimental result of which are illustrated as Fig.2,3 respectively.

%\figurename~\ref{fig:result1} and \ref{fig:result2}  illustrate 
The performance validation of various methods is based on two commonly used DNN models: ResNet101 and VGG19. The performance is evaluated from three aspects: \textit{task completion rate}, \textit{total average delay}, and the \textit{variance} in the total workload assigned to each satellite.

%\textit{task completion rate} ($1-r_D$),  \textit{total average delay} that indicates the average $t^{sum}_x$ of all satellites, \textit{resource usage variance} that shows the variance of sum of workload assigned to each satellite.

As depicted in Figs. \ref{fig:result1}(a) and \ref{fig:result2}(a),  \textit{SCC} exhibits the capability to maintain high performance, even when the task incidence is relatively high. This is attributed to the dynamic and adaptive insertion operation, along with a unique deficit calculation method employed by \textit{SCC}. These features enable \textit{SCC} to explore a vast solution space, resulting in an approximately 4\% increase in task completion rate against the others.% compared with the other methods.

%initially, the completion rate of various methods maintain a high performance.% since the drop rate is taken into the calculation of deficit %With the increase of task incidence, \textit{SCC} discriminates with other methods when $\lambda = 20$. %\note{Explain why SCC does well here}Thanks to the dynamically adaptive inserting operation and unique deficit calculation, SCC could search largely among the solution space for a better scheme. 

From the aspect of total average delay, the performance of each method increases proportionally with the growth in task incidence. Among these methods, our proposed \textit{SCC} maintains a relatively low delay performance by optimizing both transmission and computation delay. In contrast, both \textit{RRP} and \textit{DQN} prefer to select the fittest satellites, leading to an imbalanced distribution where a particular satellite is chosen by multiple decision-making satellites. As shown in Figs~\ref{fig:result1}(b) and \ref{fig:result2}(b), on average, \textit{SCC} reduces the delay by 620~ms and 140~ms against \textit{RRP} and \textit{DQN} respectively. %Therefore it is more probably for a satellite with more resource to be designated by couple of satellites simultaneously, which may cause an imbalance distribution of tasks and lead to a worse perform. 

In terms of the variance in satellite usage, a smaller value signifies a more robust consideration of load balancing across different satellites. With the assistance of balanced task splitting, our proposal effectively harnesses the available resources. \textit{SCC} can achieve a similar performance compared with \textit{Random}, which can theoretically achieve a perfectly even distribution.

%Specifically, the performance of \textit{SCC} is only inferior to \textit{Random} which could fairly implement absolutely even distribution mathematically. %but the performance of \textit{Random} in other criteria is not acceptable.

%To validate the effectiveness of our proposal across different network scales, we verify the relationship between task completion rate. 
We verify the task completion rate with different network scales. The generated task incidence is fixed as 25. %As illustrated in Figure~\ref{fig:result1}(d) and \ref{fig:result2}(d), 
The results demonstrate that \textit{SCC} can still outperform other methods even if the network scale is more than 1000 (= 32$\times$32). This is because \textit{SCC} tends to choose satellites with low deficits, indicating that the selected satellites currently possess more resources available for offloading decisions. This leads to a more balanced task distribution of efficient offloading. 

%This scheme helps in achieving a more balanced distribution of task offloading.

%takes the usage of satellite into calculating deficit. More precisely, using a satellite with less vacant resource is prone to lead to a higher deficit.

\section{Conclusion}
In this paper, we delve into the challenge of DNN task splitting and offloading within a collaborative satellite computing system with the objective of minimizing both task delay and drop rate. \note{To achieve this, we have developed an adaptive task splitting scheme designed to balance the workload efficiently. This scheme dynamically orchestrates collaborative inference among satellites, significantly enhancing the utilization rate of satellite computing resources. Furthermore, a GA-based self-adaptive task offloading scheme is introduced to determine optimal offloading decisions.} The experimental results demonstrate the superiority of our proposal over comparable methods in terms of overall delay and task completion rate. Our primary forthcoming focus revolves around the integration of an early exit technique that balances the trade-off between processing delay and accuracy during the DNN partitioning process.

\bibliographystyle{ieeetr}
{\small \bibliography{ref.bib}}

\end{CJK}
\end{document}